\documentclass[%
 reprint,
 amsmath,amssymb,
 aps,prb
]{revtex4-1}
\usepackage{graphicx}
\usepackage{dcolumn}
\usepackage{bm}

\usepackage{braket}
\usepackage{xcolor}
\begin{document}
\title{Light Induced Quantum Anomalous Hall Effect in Cubic Rashba Spin-Orbit Coupled Systems}

\author{Debabrata Sinha}
\affiliation{Department of Physics, Ramakrishna Mission Vidyamandira, Belur Math, Howrah 711 202, India}

\date{\today}

\begin{abstract}
We investigate topological phase transitions in a two-dimensional electron system with cubic Rashba spin–orbit coupling driven by circularly polarized light. Within the Floquet framework, we demonstrate that light–matter interaction induces nontrivial band topology characterized by a quantized anomalous Hall response, with Chern insulating phases of $\mathcal{C} = 0, \pm 1,$ and $\pm 3$. These transitions are governed by gap closings at high-symmetry points in the Brillouin zone, controlled by intensity and energy of the incident light. Introducing a weak linear Rashba term displaces Dirac points in momentum space without modifying the topology, whereas a purely linear Rashba system remains topologically trivial ($\mathcal{C} = 0$). When both linear and cubic Rashba couplings are finite, the linear term confines nonzero-Chern phases to narrow parameter windows. In contrast, incorporating a linear Dresselhaus term into the cubic Rashba system can trigger topological transitions even at small coupling strengths. These results clarify the interplay between distinct spin–orbit couplings in Floquet-engineered Chern insulators and offer experimentally relevant pathways for achieving light-controlled topological phases.

\end{abstract}
\maketitle
\section {Introduction}
The discovery of the quantum Hall effect established a direct link between the topological properties of a material’s bulk band structure and the presence of dissipationless edge states circulating along its boundary\cite{Zhang-Nat05},\cite{Klitzing-PRL80}. The relevant topological property is quantified by the Chern number, $\mathcal{C}$, defined as the integral of the Berry curvature over a completely filled band\cite{Thou-PRL82}, and equivalently equal to the number of chiral edge modes at the boundary. The quantum anomalous Hall effect (QAHE) originates from the same topological principle, but it occurs in the absence of an external magnetic field, relying instead on intrinsic magnetic oredering and strong spin–orbit coupling\cite{Mac-RMP},\cite{Yu-Science10},\cite{Chang-science13},\cite{Chang-Nat15},\cite{Bestwick-PRL15}. The Hall conductivity in this phase is quanized as $\mathcal{C}e^2/h$  where $h$ is the Planck constant and $e$ is the elementary charge\cite{Haldane-PRL88}. While Hall insulators with $\mathcal{C}=\pm 1$ has been extensively studied, system with large Chern numbers are of particular interest, as they can subtantially reduce contact resistance and are therefore highly desirable for device applications\cite{Zhao-Nat20},\cite{Wang-PRL13},\cite{Fang-PRL14}. Recently, high Chern number states have been reported in magnetically doped topological insulators\cite{Jiang-PRB12}, multilayer graphene\cite{Liu-PRB25}, and higher-order topological insulators\cite{Wan-PRB24}.

The spin-orbit couplings plays a key role in realizing the QAH in two dimensional electron  systems\cite{Han-Science24},\cite{Dugaev-PRB05}. SOI arises from the coupling of electron's spin with its orbital motion in the presence of an electric field, which originate mostly form from structural inversion asymmetry (Rashba-type SOI)\cite{Rashba-Sloid} or bulk inversion asymmetry (Dresselhaus-type SOI)\cite{Dress}. In low-dimensional systems, such as semiconductor quantum wells, transition metal dichalcogenides, the magnitude and symmetry of SOI can be tuned by external gates, strain, layer stacking. This tunability enables precise control over band inversion, a prerequisite for topological insulating phases.

In addition to the conventional $k$-linear Rashba and Dresselhaus couplings, a higher-order $k^3$ Rashba SOI has recently drawn growing attention, owing to its significant impact on the band structure and spin textures of low-dimensional materials\cite{Moriya-PRL14},\cite{Naka-PRL12},\cite{Ji-PRB24},\cite{Li-PRB25}. The cubic Rashba SOI has been reported to found in interface such as $(111)$ surface of inversion asymmetric semiconductor GaAs\cite{Koga-PRB02},\cite{Minkov-PRB05}, two dimensional electron gas formed at the surface of oxide SrTiO$_3$\cite{Caviglia-PRL10} and also at the oxide interface LaAlO$_3$/SrTiO$_3$\cite{Zhou-PRB15}. The coexistence of cubic and linear couplings alters the symmetry of the effective spin–orbit field and leads to notable modifications in the band structure. In the case of a pure cubic Rashba coupling, a higher-order Dirac point can emerge at a high-symmetry point in the Brillouin zone (BZ), and the presence of such a Dirac point can give rise to a higher Chern number. The presence of a linear Rashba or Dresselhaus coupling lifts this degeneracy, splitting the single higher-order Dirac point into multiple Dirac points. This splitting alters the Berry curvature distribution and consequently changes the Chern number, establishing a direct link between spin–orbit coupling symmetry and the topological characteristics of the band structure.

 However, achieving topological phase in such spin–orbit–coupled systems requires breaking time-reversal symmetry\cite{Kane-PRL05},\cite{Fu-PRB07},\cite{Ghorashi-arxiv}. A particularly versatile approach is provided by Floquet engineering, in which periodic driving—such as irradiation with circularly polarized light—effectively breaks time-reversal symmetry and opens topologically nontrivial band gaps\cite{Linder-Nat11},\cite{Gedik-Science2013}. This technique enables dynamic control over the Berry curvature distribution, allowing both the magnitude and the sign of the Chern number to be tuned in systems with strong spin–orbit coupling. In this work, we demonstrate that a quantum anomalous Hall (QAH) phase with a tunable Chern number can be engineered in a system dominated by cubic Rashba spin–orbit coupling through the application of periodic driving within the Floquet framework. Our analysis reveals that by controlling the intensity or photon energy of the incident light, the system can be driven into distinct topological phases characterized by Chern numbers $\mathcal{C} = \pm 3, \pm 1, 0$. We further explore the effect of adding linear Rashba and Dresselhaus spin–orbit couplings to the system. In isolation, the linear Rashba interaction does not induce a topological phase transition. However, when it coexists with cubic Rashba coupling, the Floquet-driven system exhibits a much richer topological structure. In this case, the accessible Chern numbers expand to $\mathcal{C} = \pm 3, \pm 2, \pm 1, 0$, achievable by tuning the light intensity. Notably, these additional nontrivial topological phases appear only within restricted parameter windows of light–matter coupling, emphasizing the delicate interplay between different spin–orbit terms and external driving. In contrast, the inclusion of linear Dresselhaus coupling alone also fails to generate a topological phase transition. Strikingly, when cubic Rashba coupling is present, the system can again host nontrivial topological phases. Unlike the Rashba case, however, here the QAH phases are found to persist across the entire range of driving parameters, highlighting a fundamental distinction between the influence of Rashba and Dresselhaus interactions in shaping Floquet topological phases.

\section{Spin-orbit coupled Hamiltonian}
The Rashba and Dresselhaus spin–orbit coupling, which are linear in the electron’s crystal momentum $k$, can be described by the following Hamiltonians\cite{Rashba-Sloid},\cite{Dress},\cite{Manchon-Nat15}:
\begin{eqnarray}
H_R=\alpha_R(k_y\sigma_x-k_x\sigma_y)
\label{rashba-Hamil}
\end{eqnarray}
for the Rashba SOI, and 
\begin{eqnarray}
H_D=\alpha_D(k_x\sigma_x-k_y\sigma_y)
\label{dress-Hamil}
\end{eqnarray}
for the Dresselhaus SOI. Here, $\alpha_R$, $\alpha_D$ denote coupling strength of the Rashba and Dresselhaus interactions, respectively, ${\bf k}=(k_x,k_y)$ is the in-plane crystal momentum of the electron, and ${\bf \sigma}=(\sigma_x,\sigma_y,\sigma_z)$ are the $2\times 2$ Pauli matrices representing the spin degree of freedom. 
In addition to the linear-in-$k$ SOI term, a cubic Rashba term can also arise in a system with strong spin-orbit coupling. The Hamiltonian is given by\cite{Naka-PRL12}:
\begin{eqnarray}
H_c=\frac{\beta}{2i}\big[k^3_-\sigma_+-k^3_+\sigma_-]
\label{cubic-rashba-Hamil}
\end{eqnarray}
where $\beta$ is the strength of the cubic Rashba coupling, $k_\pm=k_x\pm i k_y$, and $\sigma_\pm=\sigma_x\pm i\sigma_y$. This higher order term plays crucial role in determining the spin splitting and the topological characteristics of the electronic band structure. The Rashba coupling strength $\alpha_R$ typically ranges from $0.05-1$ eV${\AA}$, while the Dresselhaus coupling strength $\alpha_D$ lies between $0.1$ and $0.5$ eV${\AA}$. Both of these parameters can be tuned by applying an external gate field or by adjusting quantum well confinement. The cubic Rashba coupling $\beta$ in a 2D electron system is generally of the same order as $\alpha_R$, whereas in a 2D hole gas, it can be significantly larger\cite{Winkler}.

\begin{figure}
\includegraphics[scale=.42]{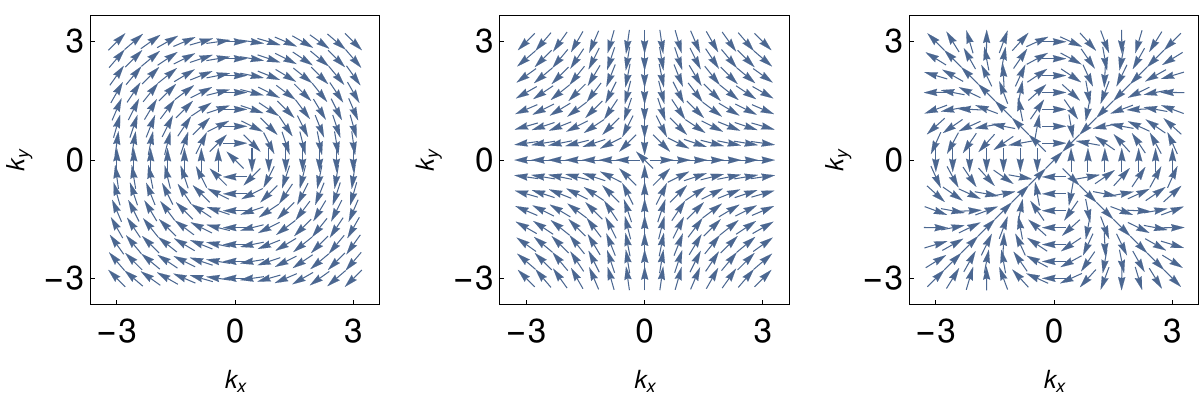}
\caption{Momentum-space spin textures for three types of spin–orbit coupling: linear Rashba (left), linear Dresselhaus (middle), and cubic Rashba (right).}
\label{spin-plot}
\end{figure}

 The spin orbit coulings in Eq.(\ref{rashba-Hamil}), (\ref{dress-Hamil}) and (\ref{cubic-rashba-Hamil}) exhibits distinct symmetry characteristics. The linear Rashba coupling breaks inversion symmetry $\mathcal{I}$, but remains invariant under the combined symmetry $\mathcal{C}_{4z}\mathcal{T}$ symmetry. Here, $\mathcal{C}_{4z}=e^{i\frac{\pi}{4}\sigma_z}$ is the fourfold rotational operator and $\mathcal{T}=-i\sigma_y\mathcal{K}$ (where $\mathcal{K}$ is the complex conjugate) is the time reversal operator. In contrast, the linear Dresselhaus coupling breaks both $\mathcal{I}$ and $\mathcal{C}_{4z}\mathcal{T}$ individually but remains invariant under the combined operator $\mathcal{I}\mathcal{C}_{4z}\mathcal{T}$. Similarly, the cubic Rashba coupling also preserves the $\mathcal{I}\mathcal{C}_{4z}\mathcal{T}$ symmetry. The distinct symmetries are evident in spin texture as shown in the Fig(\ref{spin-plot}). 

To understand the topological properties of spin texture in different spin-orbit Hamiltonian, it is necessary to analyze the spin winding number, which describe how the in-plane spin vector ${\bf S_\parallel=(S_x,S_y)}$ rotates as the momentum ${\bf k}$ varies along a closed loop in the BZ\cite{Ji-PRB24}. The spin winding number is defined via the following contour integral in the BZ:
\begin{eqnarray}
\mathcal{W}=\frac{1}{2\pi}Im \oint \frac{dS_x+i dS_y}{S_x+iS_y}
\label{wind}
\end{eqnarray}
The winding number has finite value if the integral traces a closed path around a singularity (e.g., band crossing or spin vortex) in the BZ. We analyze the spin texture arising from different form of spin orbit interactions and compute their associated winding number near the time-reversal invariant momenta (TRIMs) in the BZ of a square lattice. 

In the case of cubic Rashba coupling, the spin texture near the $\Gamma$ point (i.e., ${\bf k}=0$), is given by: ${\bf S}_{\parallel}=(-\sin 3\theta, \cos 3\theta)$, where $\theta=\arctan(k_y/k_x)$ is the polar angle in momentum space. The in- plane spin winds three times around the origin when $\theta$ varies from $0$ to $2\pi$, implies that $\mathcal{W}_{\Gamma}=3$. In contrast , for the linear Rashba coupling, the spin texture near the $\Gamma$ point takes the form: ${\bf S}_{\parallel}=(\sin \theta, -\cos \theta)$ implying the in-plane spin winds once around the origin when $\theta$ traverses $2\pi$. Consequently, the corresponding winding number is $\mathcal{W}_{\Gamma}=+1$. Similarly, for the linear Dresselhaus coupling, where the spin texture ${\bf S}_{\parallel}=(\cos\theta,-\sin\theta)$, the spin winding number is $\mathcal{W}_{\Gamma}=-1$.

According to the poincare-Hopf theorem, the total winding number over the  enire Brillouin zone must vanish, i.e.,  $\sum_i\mathcal{W}_i=0$, where the summation runs over all isolated singularities in the spin texture, typically at the TRIM points. We compute winding numbers at the remaining TRIM of a square lattice BZ, namely :  $X=(\pi,0)$, $Y=(0,\pi)$ and $M=(\pi,\pi)$. The low energy Hamiltonian around these points are linear and can be written as $H=\sum v_{ij}k_i\sigma_j$. The spin winding number associated with each TRIM is determined by the sign of the determinant of $v_{ij}$ i.e., $\mathcal{W}=\operatorname{sgn}(\det v)$. For cubic Rashba coupling, we find $\{\mathcal{W}_X,\mathcal{W}_Y,\mathcal{W}_M\}=(-1,-1,-1)$ which compensate $\mathcal{W}_{\Gamma}=+3$. The spin winding numbers at high symmetry points for linear Rashba coupling: $\{\mathcal{W}_{X},\mathcal{W}_{Y},\mathcal{W}_{M},\mathcal{W}_{\Gamma}\}=(-1,-1,+1,+1)$. Similarly, for linear Dresselhaus coupling: $\{\mathcal{W}_{X},\mathcal{W}_{Y},\mathcal{W}_{M},\mathcal{W}_{\Gamma}\}=(+1,+1,-1,-1)$.

\begin{figure}
\includegraphics[scale=.4]{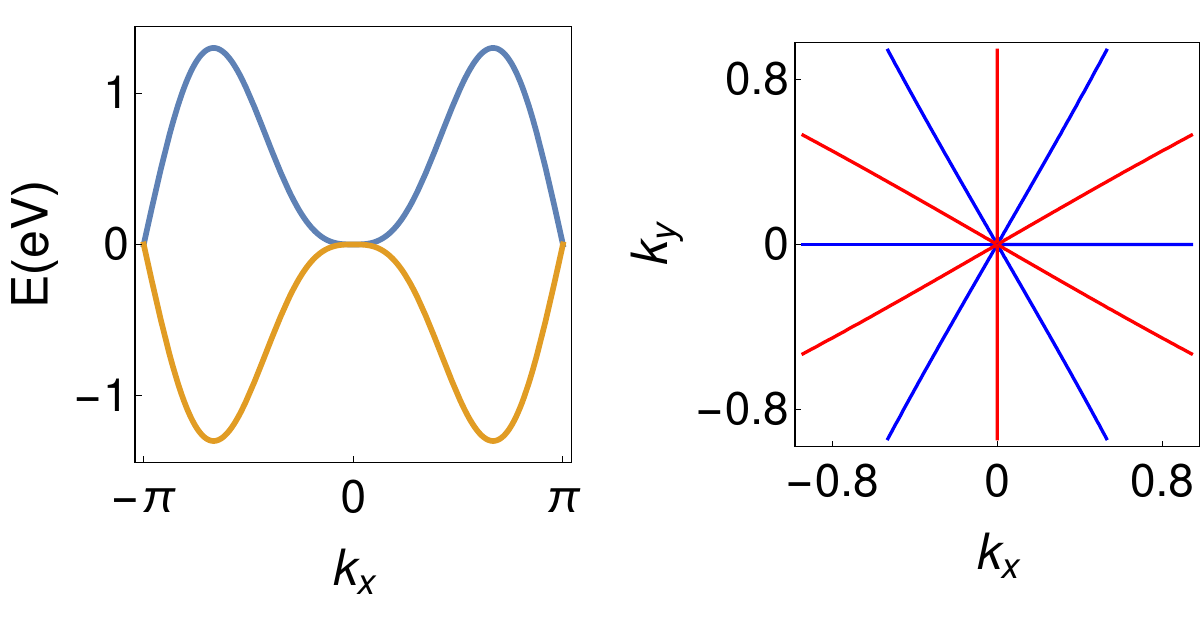}
\includegraphics[scale=.4]{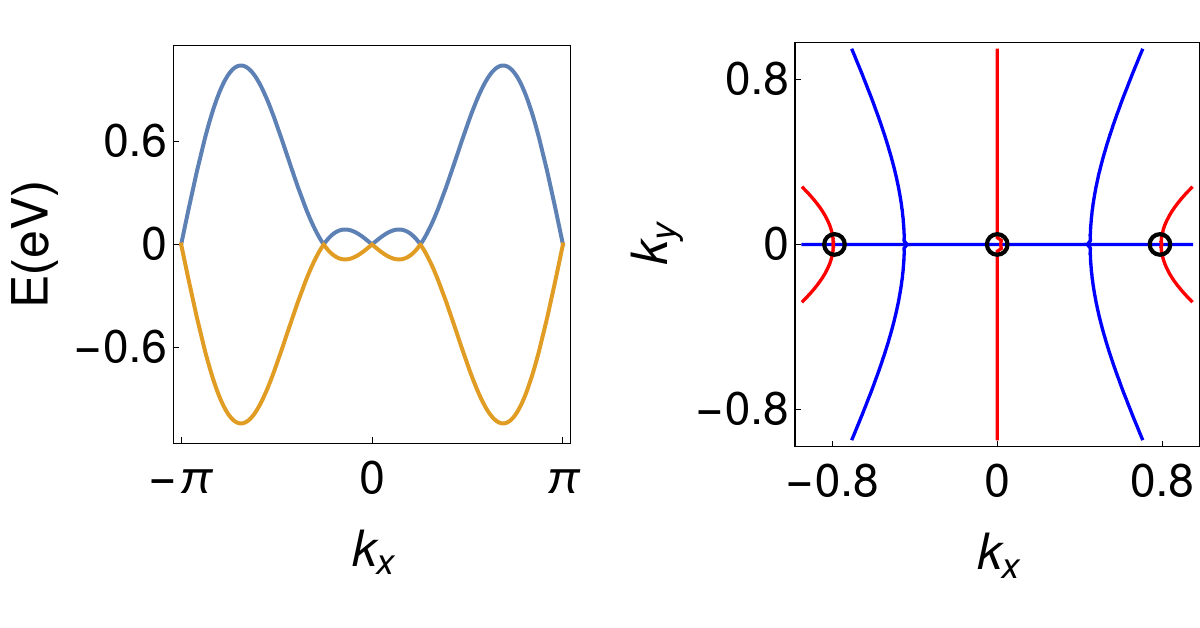}
\includegraphics[scale=.4]{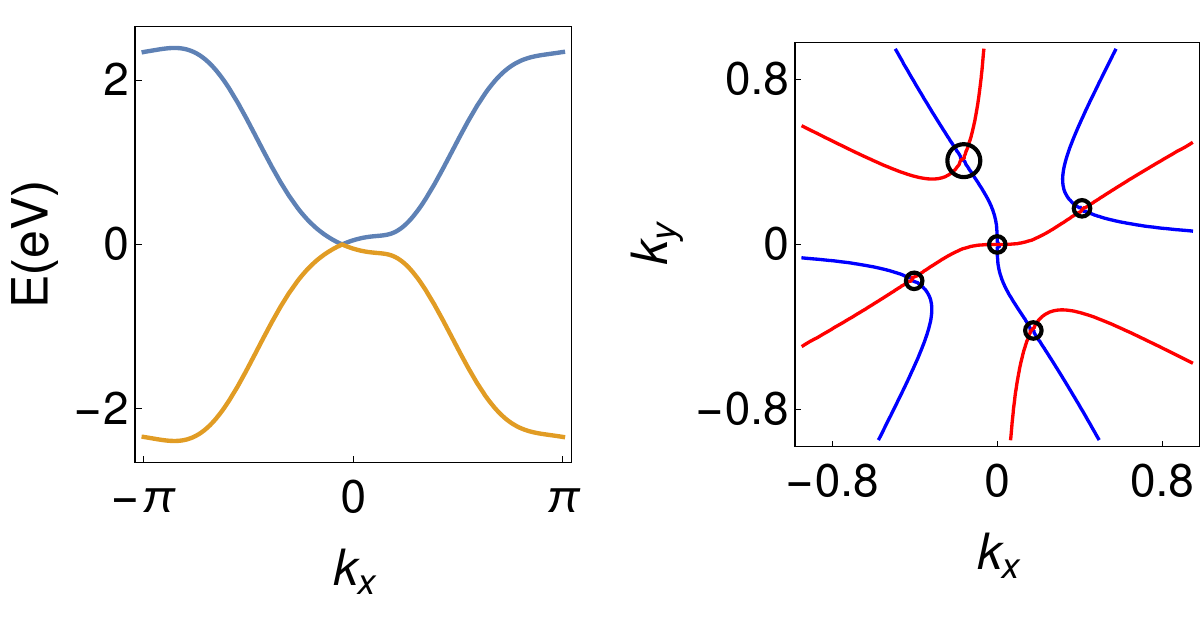}
\caption{The top panel shows the energy band structure and spin contours of the in-plane spin vector field $(S_x,S_y)$ for the case of cubic Rashba spin-orbit coupling. The blue and red lines represents the contours where $S_x=0$ and $S_y=0$, respectively. The energy dispersion is plotted as a function of $k_x$ with $k_y=0$. The band touching point at ${\bf k}=(0,0)$ indicates the presence of a nonlinear Dirac point, characteristic of cubic spin-orbit interaction. The middle panel is shown for both finite linear and cubic Rashba coupling. We fix $\alpha_R=0.3$ eV${\AA}$ and $\beta=0.5$ eV${\AA}^3$. The three black circles in spin contour plot are the position of linear Dirac points. The bottom panel is shown for both finite linear Dresselhaus and cubic Rashba coupling. We fix $\alpha_D=0.1$eV ${\AA}$ and $\beta=0.5$ eV ${\AA}^3$. The position of five Dirac points are shown in the spin contour plot by black circle. The energy diagram is shown at the point marked by a black big circle. The splitting of multiple Dirac points has also been discussed in Ref.\cite{Ji-PRB24}.}
\label{spin_eng-plot}
\end{figure}

When both linear and cubic spin-orbit coupling are present, the interplay between them modifies the band structure, splitting the higher order topological defects into multiple elementary ones. In the pure cubic Rashba model, the $\Gamma$ point hosts higher order winding number $\mathcal{W}=+3$, corresponds to a single nonlinear Dirac point. This nonlinear Dirac point is protected by the $C_{3z}$ rotational symmetry. The upper panel of Fig.(\ref{spin_eng-plot}) shows the energy band diagram and spin contour for the pure cubic Rashba coupling. The band diagram reveals a nonlinear Dirac point at ${\bf k}=(0,0)$ while in the spin contour plot, the intersection of $S_x=0$ lines (shown by blue) with $S_y=0$ lines (shown by red) identifies the Dirac pointwith $\mathcal{W}=3$\cite{Ji-PRB24}.

Introducing a linear Rashba term breaks the $C_{3z}$ symmetry, rendering the higher order vortex unstable and causing to split into three Dirac points, each with winding number $\mathcal{W}=+1$, arranged symmetrically around the origin as shown in the middle panel in Fig.(\ref{spin_eng-plot}). The location of these Dirac points can be obtained by finding the zeros of the in-plane spin-orbit field ${\bf d}_\parallel({\bf k})=(d_x,d_y)$ yields three solutions:
\begin{eqnarray}
{\bf k}_0=(0,0), \pm {\bf K}_\alpha=(2\arcsin \sqrt{\frac{\alpha_R}{4\beta}},0))
\label{new-Dirac}
\end{eqnarray}
where one Dirac point remains at the origin, while two other two appears symmetrically along the $k_x$ axis. Each is characterized by a linear band crossing and a winding number $\mathcal{W}=+1$. The total topological charge around the $\Gamma$ point is conserved i.e., $\mathcal{W}^{total}_{\Gamma}=\mathcal{W}_0+\mathcal{W}_++\mathcal{W}_-=+3$. When $\alpha_R=4\beta$, the off-center Dirac points at $\pm K_\alpha$ merge with the Dirac point located at $X$, triggering a topological phase transition. Consequently, the topological cahrge at $X$ changes from $\mathcal{W}=-1$ to $\mathcal{W}=+1$.  

The presence of linear Dresselhaus also breaks the threefold rotational symmetry that is associated with cubic Rashba coupling. The linear Dresselhaus coupling respects only twofold and mirror symmetries $C_{2v}$. As a result of this reduced symmetry and topological requirement to conserve total winding number, the original nonlinear Dirac point split into five distinct linear Dirac points as shown in the botom panel in Fig.(\ref{spin_eng-plot}). The position of the Dirac points can be obtained by numerically solving ${\bf d}_{\parallel}({\bf k})=0$.  The centre point is associated with winding number $\mathcal{W}=-1$ and remaining four points are carrying winding number $\mathcal{W}=+1$.

\section{Floquet Formalism} 
We consider the cubic Rashba Hamiltonian in Eq.(\ref{cubic-rashba-Hamil}) and subject the system to an off-resonant circularly polarized light with electric field $\vec{E}(t)=E_0(-\sin \omega t,\eta\cos\omega t)$, where $E_0$ is the amplitude of the electric field, $\eta$ is the polarization of light. The Hamiltonian of the driven system become a time-periodic i.e. $H(\vec{k},t)=H(\vec{k},t+T)$ with $T=2\pi/\omega$. The corresponding vector potential $\vec{A}(t)$ is given by,
\begin{eqnarray}
\vec{A}(t)=A_0(\cos \omega t, \eta \sin \omega t)
\end{eqnarray}
where, $A_0=E_0/\omega$. The electromagnetic field coupled with electron's momentum by Peierls substitution i.e., $\vec{k}$ in Eq.(\ref{cubic-rashba-Hamil}) is now replaced by $\vec{k}+\frac{e}{\hbar}\vec{A}$. The time dependent Hamiltonian can now read as,
 \begin{eqnarray}
 H(\vec{k},t)=H_0(k)+\mathcal{V}(\vec{k},t)+\mathcal{O}(\vec{A}^2)
 \label{td}
 \end{eqnarray}
 where, $H_0$ is the Hamiltonian of the undriven system. The static part $H_0(k)$ describes a two-dimensional electron system with cubic Rashba coupling, and is given by,
 \begin{eqnarray}
 H_0(k)&=&\frac{\hbar^2k^2}{2m}+\beta(k^3_y-3k^2_xk_y)\sigma_x\nonumber\\&&+\beta(k^3_x-3k_xk^2_y)\sigma_y+\Delta_m\sigma_z
 \label{cr}
 \end{eqnarray}
 where the first term denotes the kinetic energy, and the second and third terms represents the cubic Rashba spin-orbit interaction. The last term, proportional to $\Delta_m \sigma_z$, is a mass term that originates from the exchange interactions when the system is placed on a ferromagnetic substrate, subjected to magnetic doping, or expose to an external Zeeman field\cite{Var-PRB25}. The time-dependent part of the Hamiltonian in Eq~(\ref{td}) is given by,
 \begin{eqnarray}
 \mathcal{V}(\vec{k},t)&=&\frac{3e\beta}{\hbar}[(k^2_y-k^2_x)A_y\sigma_x-2k_xk_yA_x\sigma_x\nonumber\\&&+(k^2_x-k^2_y)A_x\sigma_y-2k_xk_yA_y\sigma_y]
 \end{eqnarray}
The time evolution of an eigenstate of the periodic Hamiltonian is given as $\ket{\Psi_k(t)}=U_{k}(t,t_0)\ket{\Psi_k(t_0)}$, where $U_k(t,t_0)=\mathcal{T}\exp[(-i/\hbar)\int^t_{t_0}H(\vec{k},t')dt']$\cite{Kitagawa-PRB11}. For an off-resonant light where the incident photon energy is much larger than the bandwidth of the system, the time evolution operator over one time period is $\mathcal{T}exp[-(i/\hbar)\int^T_0 H(\vec{k},t')dt']=exp[-(i/\hbar)H_F(\vec{k})T]$. The floquet Hamiltonian is written as\cite{Mikami-PRB16},\cite{Eckardt-NJP15},\cite{Sinha-EPL16},
\begin{eqnarray}
H_{F}(\vec{k})=H_0(\vec{k})+\sum_{n\geq 1}\frac{[\mathcal{V}_{-n},\mathcal{V}_{n}]}{n\hbar \omega}+\mathcal{O}(\frac{1}{\omega^2})
\label{Fl-an}
\end{eqnarray}
where $\mathcal{V}_{+n}=(1/T)\int^T_0 e^{-in\omega t}H(\vec{k},t)dt$. The effective floquet Hamiltonian can be recast in the general two-band form,
 \begin{eqnarray}
 H_{F}(\vec{k})=d_0(\vec{k})\sigma_0+\vec{d}(\vec{k})\cdot \vec{\sigma}
 \label{floq-hamil}
 \end{eqnarray}
 where,
 \begin{eqnarray}
 d_0(\vec{k})&=&\frac{\hbar^2\vec{k}^2}{2m}\nonumber\\
 d_1(\vec{k})&=&\beta(k^3_y-3k^2_xk_y)\nonumber\\
 d_2(\vec{k})&=&\beta(k^3_x-3k^2_yk_x)\nonumber\\
 d_3(\vec{k})&=&J_2(k^2_y+k^2_x)^2+\Delta_m
 \label{CR}
 \end{eqnarray}
where, $J_2=\eta\frac{9\beta^2 e^2A^2_0}{\hbar^3 \omega}$. The term $\Delta_m$ in Eq.(\ref{CR}) represents an exchange-induced mass term that breaks time reversal symmetry and opens a finite gap at the $\Gamma$ point, thereby making the Chern number well defined. Physically, $\Delta_m$ arises from a ferromagnetic exchange field that couples to the spin degree of freedom of two-dimensional electron gas. Such exchange field can originate from various mechanisms, including proximity coupling to ferromagnetic substrate or doping with magnetic ions or the application of an external Zeeman field.

The floquet analysis in Eq.(\ref{Fl-an}) is valid in the off-resonant regime, characterized by the condition that the dimensionless parameter $\tilde{A}_0=\frac{eA_0 a}{\hbar} \ll 1$. In this regime, the light is far detuned from electronic transitions, so it does not excite electrons across the bands and hence no real optical absorption occurs. Instead, the effect of the light is to modify the band structure through virtual photon emission and absorption process\cite{Kitagawa-PRB11},\cite{Hubner-Nat17}(see Appendix-A). For experimental relevence we set driving frequency $\omega=1000$THz. The corresponding light intensity is given by,
\begin{eqnarray}
I=\frac{c\epsilon_0(\hbar \omega)^2\tilde{A}^2_0}{2e^2a^2}
\end{eqnarray} 
where $\omega$ is the driving frequency, $c$ is the speed of light, and $\epsilon_0$ is the vacuum permittivity and $a$ is the lattice constant. For typical value of $\tilde{A}_0\sim 0.01-0.1$ and a lattice constant $a=1{\AA}$, the corresponding light intensity lies in the range $I\sim 10^{12}-10^{14}$ W/m$^2$, with an electric field amplitude $E_0\sim 10^7-10^8$ V/m. These field strength are well within the reach of current pump–probe and Floquet engineering experiments\cite{Gedik-Science2013},\cite{Gedik-Na16}. The frequency of the dressing field (1000 Thz $\sim$ 4eV) ensures operation in the off-resonant regime relative to the typical bandwidths of semiconductors ($\sim$ 0.5-2 eV).

A periodic light field continuously injects energy into the electronic system. For Floquet-engineered topological phases to be experimentally observable, this injected energy must be balanced by dissipation into the environment through electron–phonon interactions, impurity scattering, or substrate coupling. In the off-resonant regime considered here, real photon absorption is strongly suppressed, and the light primarily induces virtual band renormalization within the coherent dynamics described by the high-frequency expansion. Under realistic conditions, the small residual energy absorbed by the electrons can be efficiently dissipated through standard relaxation processes, leading to effective damping of the electronic motion. This balance between energy absorption and dissipation enables a quasi-steady Floquet topological state to persist over experimentally relevant timescales. A qualitative discussion of this energy balance and damping mechanism is provided in Appendix B, where it is emphasized that these considerations serve only as experimental context and do not alter the formal isolated-system framework used in our analysis.

To investigate the topological properties numerically, we discretize the model on a square lattice. The corresponding lattice Hamiltonian reads (with lattice constant $a=1 {\AA}$),
\begin{eqnarray}
H_{lattice}&=&2t(\cos k_x+\cos k_y)\sigma_0\nonumber\\&&+4J_2(2-\cos k_x-\cos k_y)^2\sigma_z\nonumber\\&&-2\beta(2-3\cos k_x+\cos k_y)\sin k_y\sigma_x\nonumber\\&&-2\beta(2-3\cos k_y+\cos k_x)\sin k_x\sigma_y+\Delta_m \sigma_z\nonumber\\
\label{lattice-hamil}
\end{eqnarray} 
This lattice regularization allows direct computation of topological invariants such as Chern number as well as numerical simulation of edge states in a finite geometry.

The Berry curvature for a two band Hamiltonian can be computed using the following expressions,
 \begin{eqnarray}
 \Omega_\pm=\mp \frac{1}{2|\vec{d}|^3}\hat{d}\cdot(\partial_{k_x}\hat{d}\times \partial_{k_y}\hat{d})
 \end{eqnarray}
 where $\hat{d}(\vec{k})=\vec{d}(\vec{k})/d(\vec{k})$ is the normalized $\vec{d}$ vector. The $\pm$ indices corrrespond to the upper and lowe energy bands, respectively. The Berry curvature $\Omega^n_k$ plays a central role in the determining the anomalous Hall response of the system. In particular, the intrinsic contribution to AHC at finite temperature is given by,
 \begin{eqnarray}
 \sigma_{xy}=\frac{e^2}{h}\sum_n\int \frac{d^2k}{(2\pi)^2}f^n_k \Omega^n_k
 \label{an}
 \end{eqnarray}
 where the sum runs over the occupied bands $n$, $f^n_k$ is the Fermi-Dirac distribution function, and the integration is performed over the entire first Brillouin zone. The validity of the Fermi-Dirac distribution in the Floquet system is discussed in Appendix-C. At zero temperature, the Fermi-Dirac distribution function becomes a step function and AHC simplifies to a qunatized value: $\sigma_{xy}=\mathcal{C}\frac{e^2}{h}$ where $\mathcal{C}$ is the chern number of the occupied bands. The chern number is a topological invariant and defined as the integration of Berry curvature over the Brillouin zone,
 \begin{eqnarray}
 \mathcal{C}=\frac{1}{2\pi}\oint \Omega_-d^2k
 \end{eqnarray}
 Here $\Omega_-({\bf k})$ denotes the Berry curvature of the lower occupied bands.


\section{Results and Discussions} 

We compute the Chern number of effective lattice Hamiltonian in Eq.~(\ref{lattice-hamil}). The condition for the band gap to close is obtained by solving $d_3({\bf k})=0$, which gives the following condition in,
 \begin{eqnarray}
 \Delta_m=-4J_2(2-\cos k_x-\cos k_y)^2
 \end{eqnarray}
 
 This equation specifies the points in momentum space where the conduction band and valance band touch, as well as parameter values at which such touching occurs. The gap closes at the $\Gamma$ point ($0,0$) if $\Delta_m$ vanishes. At the $M$ point ($\pi,\pi$), the gap closes when $J_2 = -\frac{\Delta_m}{64}$. Similarly, at the $X$ ($\pi,0$) and $Y$ ($0,\pi$) points, the gap closes when $J_2 = -\frac{\Delta_m}{16}$. From the gap closing conditions, the Chern number changes as a function of light-matter coupling strength $A_L$ (with $\eta=-1$):
 \begin{equation}
 \mathcal{C}= 
 \begin{cases}
 0, & 0<A_L<\frac{|\Delta_m|}{576\beta^2}, \\
 1, & \frac{|\Delta_m|}{576\beta^2}<A_L<\frac{|\Delta_m|}{144\beta^2}, \\
 3, & A_L>\frac{|\Delta_m|}{144\beta^2}.
 \end{cases}
 \label{chern-cubic}
 \end{equation}
where, $A_L=\frac{e^2A^2_0}{\hbar^3\omega}$. Thus by tuning $A_L$, the system exhibits the sequence of topological transitions $\mathcal{C}: 0\to 1 \to 3$, which in turn can be controlled by the intensity or photon energy of the incident light. The direct light-matter control offer a feasible route to engineer the system's topological properties in situ.  The AHC  exhibits corresponding quantized plateau $3e^2/h$ to $e^2/h$ to $0$, reflecting the change in the underlying topological invariant. The appearance of each plateau marks a separate topological phase, where the associated Chern number encodes the bulk band topology.
\begin{figure}
\includegraphics[scale=.5]{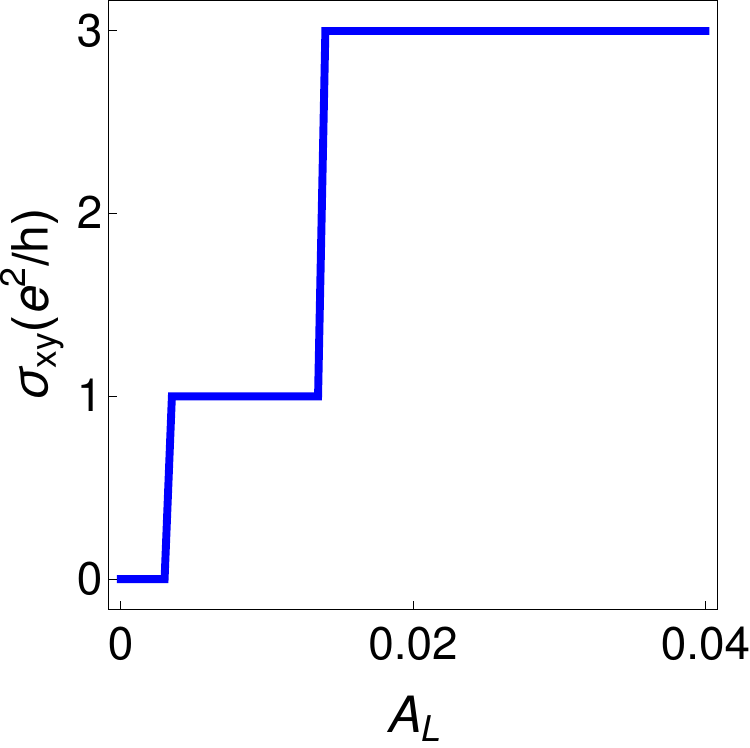}
\caption{The anomalous hall conductivity $\sigma_{xy}$ is plotted as a function of $A_L$ for the case of cubic Rashba spin-orbit coupling. Here, we fix the parameters $\beta=0.5$ eV$\cdot$${\AA}^3$, and $\Delta_m=0.5$ eV.}
\label{cubic-topo}
\end{figure}
A notable feature of this transition sequence is the absence of an intermediate $\mathcal{C}=2$ plateau. This occurs because the gap closing at the $X$ and $Y$ points of the BZ takes place simultaneously. This behavior is dictated by the underlying crystal symmetry, which ensure that these two points are energetically and topologically equivalent. As a result, the Chern number changes by two units in a single transition, producing a direct jump from $\mathcal{C}=1$ to $\mathcal{C}=3$ (or vice versa) without passing through a $\mathcal{C}=2$ phase. In our calculation, we illustrate the variation of $\sigma_{xy}$ in Fig.(\ref{cubic-topo}) with light matter coupling parameter $A_L$ for $\beta=0.5$ eV ${\AA}^3$, using left-handed circularly polarized light. Note that, reversing the light's helicity to right handed ($\eta=+1$), together with changing the sign of the ferromagnetic mass term $\Delta_m$, leads to complete inversion of of the topological invariant. Under these conditions, the sequence of Chern number with increasing the parameter $A_L$ changes from $0$ to $-1$ to $-3$. This inversion arises because both the light helicity and the sign of $\Delta_m$ determines the the direction of Berry curvature in momentum space, effectively reversing the topological bands. Importantly, the system remains topological nontrivial region if $A_L<0$ (or $A_L>0$) depending on the polarization of light and sign of $\Delta_m$.
\begin{figure}
\includegraphics[scale=.3]{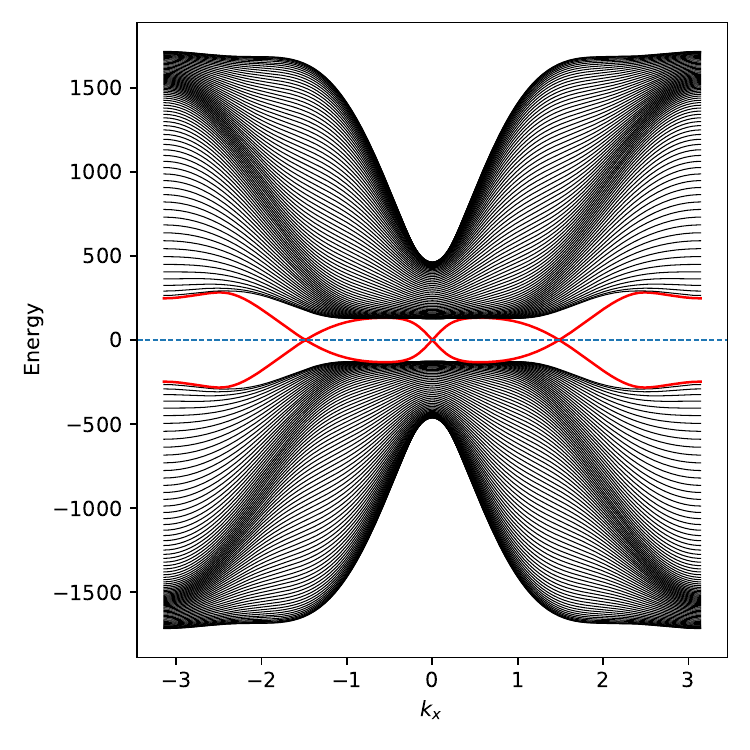}
\includegraphics[scale=.35]{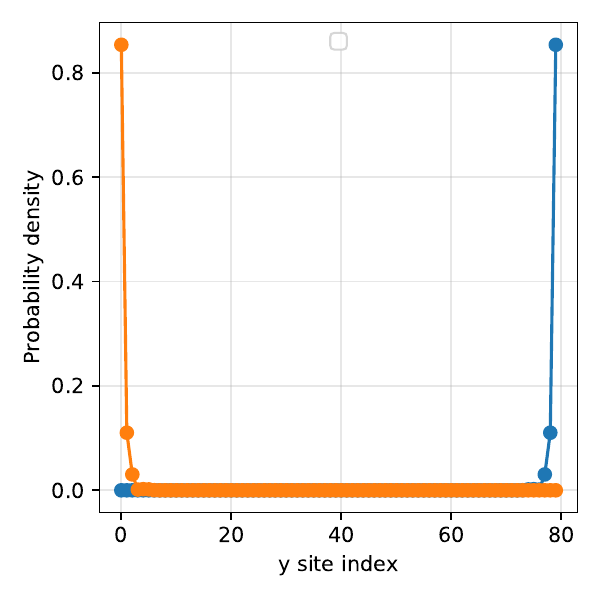}
\caption{The left panel shows the energy spectrum with $k_x$ under open boundary conditions with parameters $\alpha_R=0.1$ eV${\AA}$, $\beta=0.5$ eV ${\AA}^3$ and $A_L=0.04$ eV$^{-1}$${\AA}^{-2}$. The red curves denote the edge states within the bulk band gap. The right panel displays the corresponding probability density profiles of these edge modes.}
\label{OBC-rashba}
\end{figure}
We find that at driving frequency of $\omega=1000$THz, Eq.~(\ref{chern-cubic}) yields Chern number of unity when $\tilde{A}_0$ lies in the range $0.05-0.096$, while it increases to $3$ when $\tilde{A}_0>.096$. These values correspond to an electric field amplitude of order $E_0\sim 10^8 V/m$ which is well within experimentally accessible regime\cite{Gedik-Science2013}.

Next, we examine the effect of linear Rashba spin-orbit coupling as a perturbation to the system. In this case, the components of floquet Hamiltonian in Eq.(\ref{floq-hamil}) are given by,
\begin{eqnarray}
d_0(\vec{k})&=&\frac{\hbar^2\vec{k}^2}{2m}\nonumber\\
d_1(\vec{k})&=&\alpha_Rk_y+\beta(k^3_y-3k^2_xk_y)\nonumber\\
d_2(\vec{k})&=&-\alpha_Rk_x+\beta(k^3_x-3k^2_yk_x)\nonumber\\
d_3({\bf k})&=&[J_1(k^2_y-k^2_x)+J_2(k^2_y+k^2_x)^2+\Delta_R+\Delta_m]\sigma_z\nonumber\\
\label{linear-cubic-hamil}
\end{eqnarray}
Here, $\Delta_R=\eta\frac{\alpha^2_Re^2A^2_0}{\hbar^3 \omega}$ is a light-induced mass term, and  $J_1=\eta\frac{6\alpha_R\beta e^2A^2_0}{\hbar^3 \omega}$ is the coefficient of the anisotropic quadratic term. Crucially, all contribution to $d_3({\bf k})$ enter with same sign, implies that the Rashba terms do not by themselves induce band inversion. Therefore, a ferromagnetic mass term $\Delta_m$ is still required to drive topological phase transition. We simulate the Hamiltonian on square lattice, where the term $J_1$ breaks the the symmetry between the $X$ and $Y$ points. However, for small value of coupling strength $\alpha_R$, both the $\Delta_R$ and $J_1$ terms provides only minor corrections to $d_3({\bf k})$, since $\Delta_R/J_2<<1$ and $J_1/J_2<1$. Thus, the dominant control of the topological phase boundary remains interplay between $\Delta_m$ and the coefficient $J_2$. This indicates that the nontrivial phases with $\mathcal{C}=1$ or $\mathcal{C}=3$ remain robust against weak linear Rashba perturbation. In Fig.~(\ref{OBC-rashba}), we present the energy band structure of the lattice model with open boundary conditions along the $y$-direction, taking $\alpha = 0.1$ eV${\AA}$ and $\beta = 0.5$ eV${\AA}^3$. The left panel shows the presence of three chiral edge modes that connect the conduction and valence bands, indicating the non-trivial topological nature of the system. The right panel depicts the corresponding probability density of these edge states, which is strongly localized at the boundary, thereby confirming their edge character. The existence of such edge modes provides clear evidence of the bulk–boundary correspondence and further demonstrates that the topological invariant remains robust against small values of $\alpha$.

In contrast, when only the linear Rashba coupling is present (i.e., in the absence of cubic Rashba terms), the coefficients $J_1$ and $J_2$ in Eq.(\ref{linear-cubic-hamil}) vanish entirely. Under such condition the energy gap closes when $\Delta_m=-\Delta_R$, which leads to simultaneous gap closing at all high-symmetry points in the BZ- namely, the $\Gamma$, $X$, $Y$, and $M$ points. As a result, the system enters a topologically trivial phase characterized by Chern number $\mathcal{C}=0$. Thus cubic Rashba coupling is essential for stabilizing topologically nontrivial phases in this floquet system.

\begin{figure}
\includegraphics[scale=.22]{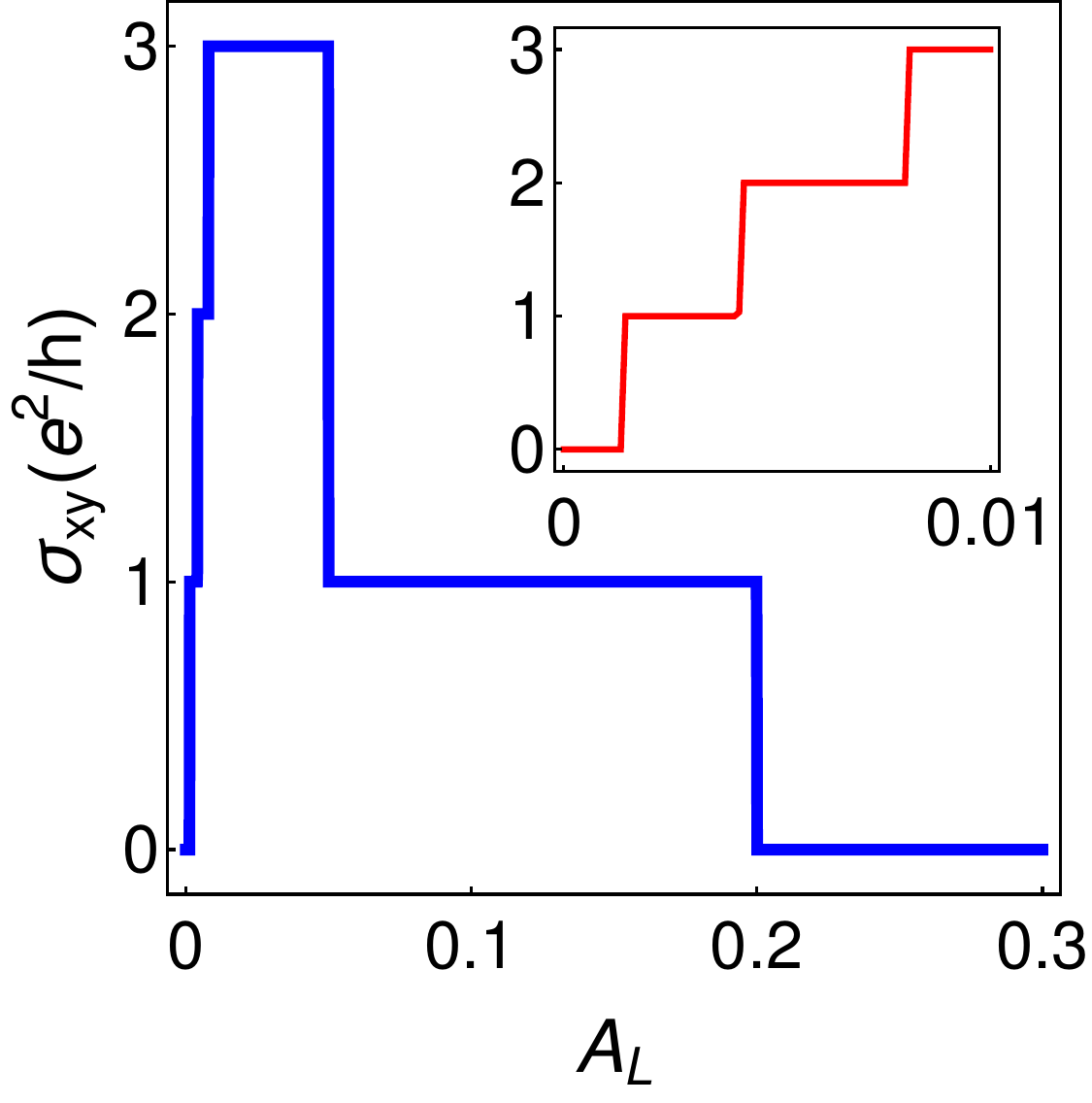}
\includegraphics[scale=.32]{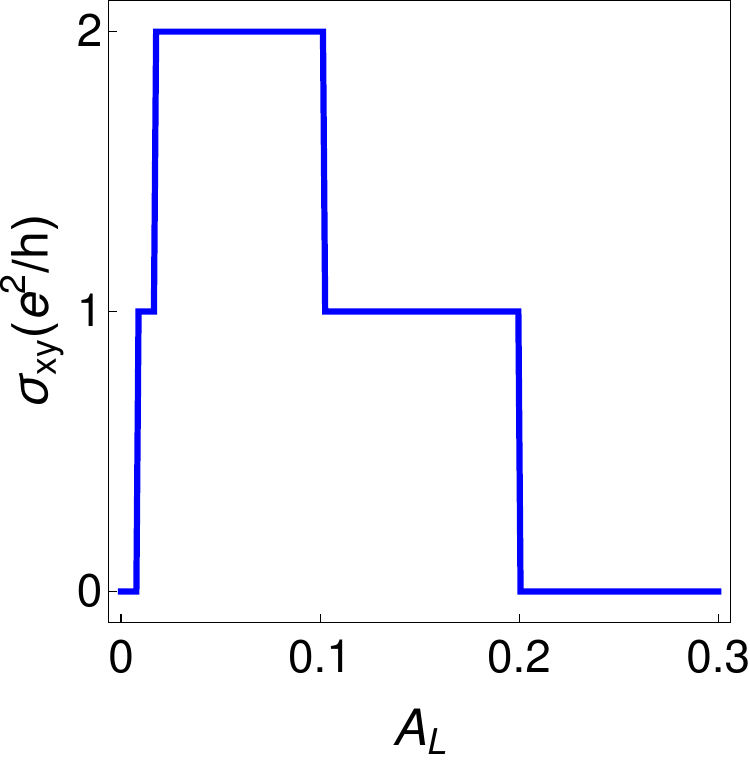}
\caption{Anomalous Hall conductivity as a function of the light coupling parameter $A_L$ at zero temperature for finite linear and cubic spin-orbit coupling strength. In the left panel, the parameters are fixed at $\alpha_R=1$ eV ${\AA}$, $\beta=0.5$ eV ${\AA}^3$. In the right panel, we set $\alpha_R=1$ eV ${\AA}$ and $\beta=0.2$ eV ${\AA}^3$, satisfying $\alpha_R> 4\beta$. In both panel $\Delta_m=0.2$ eV. }
\label{finite-rashba}
\end{figure}

We now turn to the topological phase transitions in the presence of both linear and cubic Rashba spin-orbit coupling. When the linear Rashba coupling strength $\alpha$ is finite, it lift the spin degeneracies at $\Gamma$ point in the BZ and leads to the splitting of Dirac points (see the middle panel of Fig.(\ref{spin_eng-plot})). Each of these Dirac points are associated with a distinct condition for the gap closing which are given by (for $\eta=-1$), 
\begin{equation}
 A_L= 
 \begin{cases}
 \frac{\Delta_m}{\alpha^2_R}, & \Gamma, \\
 \frac{\Delta_m}{\alpha^2_R+144\beta^2- 24\alpha_R\beta}, & X, \\
 \frac{\Delta_m}{\alpha^2_R+144\beta^2+ 24\alpha_R\beta}, & Y, \\
 \frac{\Delta_m}{\alpha^2_R+576\beta^2}, & M.
 \end{cases}
 \label{cubic-linear}
 \end{equation}
 For $\alpha<<\beta$, the contribution from $J_1$ is negligible and restoring the degeneracy between $X$ and $Y$. Finally, the shifted Dirac point $\pm{\bf K}_\alpha$ in Eq.(\ref{new-Dirac}), the gap closing depends nontrivially on $\alpha$ through the momentum displacement $K_\alpha$. Thus the interplay between $\alpha$ and $\beta$ leads to much richer phase diagram. The intermediate Chern number $\mathcal{C}=2$ become accessible and this phase emerges when gap closing at $X$ and $Y$ occurs at different critical values of tuning parameters such as light intensity or energy. The Chern number evolves as
\begin{eqnarray}
 \mathcal{C}: 0 \to 1 \to 2 \to 3 \to 1 \to 0
 \end{eqnarray}
 as the band gap sequentially closes at the points $M \to $Y$ \to {X} \to K_\alpha \to \Gamma$. Note that the Chern number change $\mathcal{C}: 3 \to 1$ here is associated with gap closing at $\pm K_\alpha$ which occurs simultaneously. Thus the presence of both $\alpha$ and $\beta$ leads to rich structure of topological transition, where the gap closing depends on the value of $\alpha$ and $\beta$. This rich topological structure manifests itself in the anomalous Hall conductivity, which shows quantized plateaus corresponds to the distinct topological phases. In particular, the variation  of AHC as a function of $A_L$- is shown in the Fig.(\ref{finite-rashba}). The figure clearly demonstrates that, a sequence of quantized transition in the Hall conductivity, corresponds to changes in the Chern number driven by gap closing at different point in the BZ.

Importantly, the inclusion of a finite linear Rashba coupling fundamentally alters the phase diagram by introducing bounded parameter windows in which topologically nontrivial phases can exist. Specifically, when both linear ($\alpha_R$) and cubic ($\beta$) Rashba coupling are present, the emergence of nonzero chern number is restricted to a finite interval of the parameter,
\begin{eqnarray}
A_L \in [0,\frac{\Delta_m}{\alpha^2_R}]
\end{eqnarray}
Beyond this interval, the system reverts to a topologically trivial phase. This behavior stands in sharp contrast to the case of pure cubic Rashba system($\alpha_R=0$), where the nontrivial phase persists over the entire range of $A_L>0$ (or $A_L<0$), without any upper bound on the light intensity or energy for maintaining the topological phase. 

This effect is illustrated in the left panel of Fig.(\ref{finite-rashba}), where we plot the variation of $\sigma_{xy}$ as a function of $A_L$ for $\Delta_m=0.2$ eV, $\alpha=1$ eV ${\AA}$ and $\beta=0.5$ eV ${\AA}^3$. The figure clearly shows that $\sigma_{xy}$ attains quantized value in the range $A_L\in [0,0.2]$ corresponding to the topologically nontrivial phase, while outside this range the quantized response vanishes, signaling a transition to a trivial insulating state. We find the maximum value of $A_L$ for the chosen parameters in Fig.(\ref{finite-rashba}), corresponding to $\tilde{A}_0=0.14$ for $\hbar \omega =0.1$ eV. The corresponding light intensity is $I \sim 2.6 \times 10^{13}$ W/m$^2$, with an electric field amplitude $E_0 \sim 1.4\times 10^8$ V/m. In the right panel, we present the variation of Hall conductivity $\sigma_{xy}$ for the parameter regime satisfying $\alpha_R>4\beta$. As discussed earlier, in this regime the band structure hosts four distinct Dirac point located at high-symmetry points $\Gamma$, $X$, $Y$ and $M$ of the BZ. The corresponding winding number associated with these points are $\mathcal{W}_\Gamma=+1$, $\mathcal{W}_X=+1$, $\mathcal{W}_Y=-1$, and $\mathcal{W}_M=-1$. The Chern number evolves with $A_L$ according to 
\begin{eqnarray}
\mathcal{C}: 0 \to 1 \to 2 \to 1 \to 0
\end{eqnarray}
as the band gap closes at the point $M \to Y \to X \to \Gamma$, respectively.

\begin{figure}
\includegraphics[scale=.35]{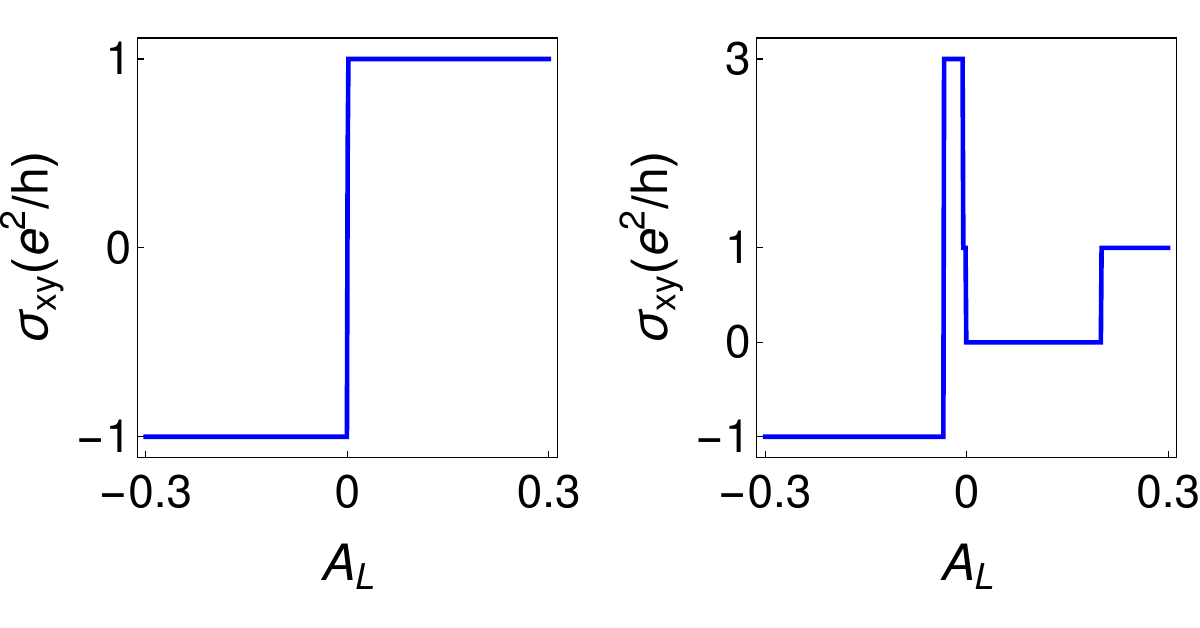}
\caption{Anomalous Hall conductivity with light coupling parameter $A_L$. In the left panel, the parameters are $\alpha_D=0.1$eV ${\AA}$, $\beta=0.5$eV ${\AA}^3$, $\Delta_m=0$. In the right panel, the parameters are  $\alpha_D=1$eV ${\AA}$, $\beta=0.5$eV ${\AA}^3$, $\Delta_m=0.1$eV.} 
\label{Dress}
\end{figure}

We now investigate the effect of linear Dresselhaus coupling, focusing on its influence on the topological properties of the cubic Rashba coupling system. The $d_3$ term in the Hamiltonian takes the following form:
\begin{eqnarray}
d_0(\vec{k})&=&\frac{\hbar^2\vec{k}^2}{2m}\nonumber\\
d_1(\vec{k})&=&\alpha_Dk_x+\beta(k^3_y-3k^2_xk_y)\nonumber\\
d_2(\vec{k})&=&-\alpha_Dk_y+\beta(k^3_x-3k^2_yk_x)\nonumber\\
d_3({\bf k})&=&-\Delta_D+4J_2(k^2_y+k^2_x)^2+\Delta_m
\end{eqnarray}
where $\Delta_D=\eta \frac{\alpha^2_De^2A^2_0}{\hbar^3\omega}$. A crucial distinction arises when we compare this scenario with the case involving linear Rashba coupling. In the Rashba case all the term in $d_3$ appear with the same sign. This prevents the natural closing of the gap, necessitating the introduction of ferromagnetic mass term to achieve a topological phase transition. However, for the Dresselhaus case, the situation is qualitatively different. The gap $\Delta_D$ and momentum dependent term to $J_2$ enter the expression for $d_3({\bf k})$ with opposite sign. The gap closing condition in the lattice model becomes (in absence of $\Delta_m$),
\begin{eqnarray}
\Delta_D=4J_2(2-\cos k_x-\cos k_y)^2
\label{gap-dress}
\end{eqnarray}
This relation shows that, for $A_L = 0$, all the high-symmetry points in the Brillouin zone become gapless. In contrast, for a finite value of $A_L$, the Chern number takes values of $\pm 1$, determined by the light polarization $\eta$. The upper-left panel of Fig.~(\ref{Dress}) illustrates this behavior. The anomalous Hall conductivity (AHC) attains the quantized value $\pm e^2/h$, and the nontrivial topological phase persists for any finite $A_L$.

In the presence of Ferromagnetic mass term $\Delta_m$, the gap closing condition is modified to
\begin{eqnarray}
 -\Delta_D+4J_2(2-\cos k_x-\cos k_y)^2+\Delta_m=0
\end{eqnarray}
which gives the condition for gap closing points at the high symmetry points in the BZ. The gapless conditions for different points are given by,
\begin{equation}
 A_L= 
 \begin{cases}
 \frac{\Delta_m}{\alpha^2_D}, & \Gamma, \\
 \frac{\Delta_m}{\alpha^2_D-144\beta^2}, & X, Y, \\
 \frac{\Delta_m}{\alpha^2_D-576\beta^2}, & M.
 \end{cases}
 \label{cubic-dress}
 \end{equation}
and also for the shifted Dirac points (see Fig.(\ref{spin_eng-plot})). From the above analysis, the Chern number is found to take the discrete values $\mathcal{C} = 0, \pm 1, 3$, as displayed in the upper-right panel of Fig.~(\ref{Dress}). This is in sharp contrast to the case of linear Rashba coupling, where a topological phase exists only within a limited window of $A_L$. We estimate the values of $\tilde{A}_0=0.14, 0.024$ and $0.012$ for the given parameters as in Fig.(\ref{Dress}) with $\hbar \omega=0.1$ eV. These values correspond to electric field amplitudes $E_0 \sim 10^7-10^8$ V/m and light intensity $I\sim 10^{11}- 10^{13}$ W/m$^2$.

\begin{figure}
\includegraphics[scale=.35]{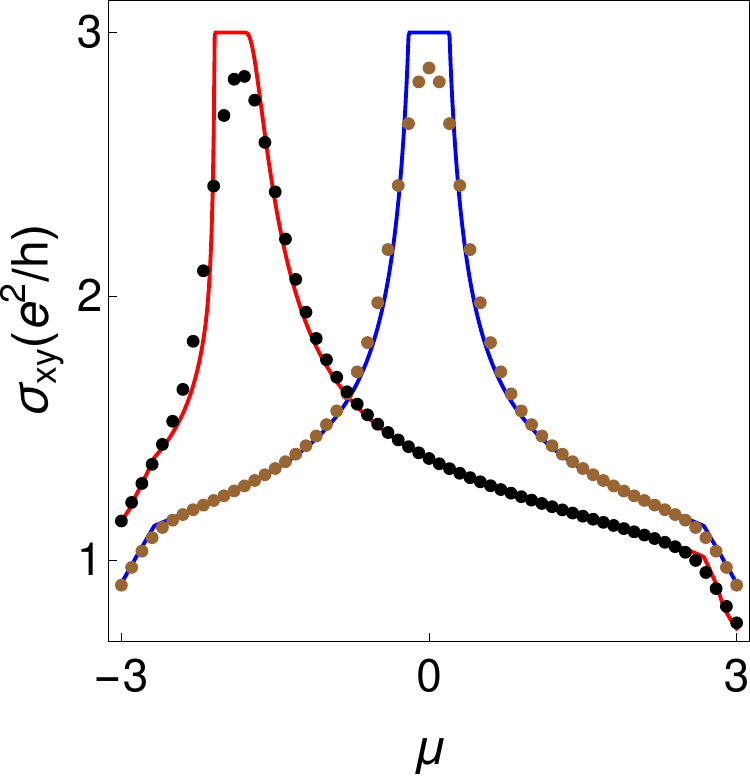}
\caption{The anomalous hall conductivity $\sigma_{xy}$, is plotted as a function of chemical potential $\mu$. The parameters are fixed as $\beta=1$ eV$\cdot$${\AA}^3$, $A_L=0.02 eV$, $\Delta_m=0.2$ eV and $\eta=-1$. The blue and red solid lines correspond to $t=0$ and $t=0.5$ eV, respectively, at $k_BT= 0.001$ eV. The dotted lines represent the results for $k_BT=0.1$ eV. }
\label{cubic-mu}
\end{figure}

Finally, we examine the influence of temperature and the kinetic energy contribution on the anomalous Hall conductivity. The inclusion of a finite $t$ (i.e., the kinetic term $\frac{\hbar^2}{2m}$) in the Hamiltonian given in Eq.(\ref{lattice-hamil}) breaks the particle- hole symmetry. As a consequence, the quantized plateau of $\sigma_{xy}$ appears at different values of the chemical potential $\mu$. In contrast, for $t=0$, the particle-hole symmetry is preserved and the plateau occurs symmetrically around $\mu=0$. The effect of temperature is to smear these quantized plateau. Fig.(\ref{cubic-mu}) shows the variation of $\sigma_{xy}$ with $\mu$ for different values of $t$ and $k_B T$. At low temperature (solid linea), $\sigma_{xy}$ remains quantized at $3e^2/h$ when $\mu$ lies in the band gap. This quantization reflects the full contribution of the Berry curvature from completely filled lower band, which carries Chern number $\mathcal{C}=3$. When $\mu$ lies deep within the lower band, incomplete occupation near the band edge reduces $\sigma_{xy}$ from its quantized value. As $\mu$ enters the upper band, opposite-sign Berry curvature from upper band partially cancels the contribution from lower band, further lowering $\sigma_{xy}$. At finite temperature (dotted line), thermal excitation cause partial occupation of states across both lower and upper bands. This lead to an incomplete cancellation of Berry curvature contribution between the two bands, which smooth out the sharp features and smears the quantized plateau in $\sigma_{xy}$.

\section{Conclusions}
 We have investigated topological phase transitions in a two-dimensional electron system with cubic Rashba spin–orbit coupling under periodic driving by circularly polarized light. Using Floquet theory, we have demonstrated that light can induce Chern insulating phases with $\mathcal{C}=0$, $1$, and $3$, with the transitions governed by gap closings at high-symmetry points in the Brillouin zone. Interestingly, the system remains in a topological phase for any finite value of light intensity or photon energy. These phases also remain robust in the presence of a weak linear Rashba coupling. However, when the linear Rashba coupling becomes finite, its coexistence with the cubic Rashba term generates a richer variety of topological phases, while confining these nontrivial phases to relatively narrow parameter ranges. In contrast, the topological phases of a pure cubic Rashba system are not robust against even a weak linear Dresselhaus coupling. Nevertheless, when a finite Dresselhaus term coexists with the cubic Rashba coupling, the system once again remains topological for any finite light intensity or photon energy. 
 
We also note that the high-Chern-number phases arise from band inversions and gap closings at the high-symmetry points $X$, $Y$, and $M$ of the Brillouin zone in Eqs. (\ref{chern-cubic}), (\ref{cubic-linear}), and (\ref{cubic-dress}). Similar momentum-space structures can naturally emerge in a wide class of artificial lattice systems and engineered two-dimensional platforms. In such systems—such as cold-atom optical lattices, photonic crystals, and oxide heterostructures—the effective Hamiltonian can remain valid across the entire Brillouin zone, allowing multiple valleys to contribute actively to the topological band structure\cite{Skirlo-PRL15},\cite{Mac-Nat17}. Additionally, high-Chern-number phases can also be realized near the $\Gamma$ point due to the splitting and hybridization of multiple Dirac points, which can drive transitions to higher-Chern-number states under suitable tuning of system parameters as shown in Fig.(\ref{gamma}). Together, these mechanisms demonstrate that our theoretical framework captures a broad range of symmetry-driven topological phases that are potentially realizable in both conventional semiconductor systems and engineered quantum platforms.

Our findings provides a deeper understanding of how distinct spin–orbit couplings shape Floquet-engineered topological phases. In particular, the ability to control Chern numbers and band topology via light–matter interaction suggests pathways for developing ultrafast, optically reconfigurable electronic devices. Potential applications include topological transistors, low-dissipation interconnects, and robust quantum channels for spintronic and photonic technologies. By tailoring spin–orbit couplings and optical driving parameters, our work provides an experimentally relevant framework for realizing light-tunable Chern insulators in quantum materials.

\appendix
\section{Floquet Analysis}
When a periodic electromagnetic field is applied to a quantum system, the Hamiltonian $H_S$ becomes time periodic, $H_S(k,t+T)=H_S(k,t)$, where $T=2\pi/\omega$ is the period of the driving field with frequency $\omega$. The dynamics of this periodic system can be described by Floquet formalism which is the extension of Bloch theorem in time domain. The time evolution of a quantum state is governed by the time-dependent Schr\text{\"o}dinger equation
\begin{eqnarray} 
i\hbar {\partial \ket{\Psi_k(t)}\over \partial t}=H_S(k,t)\ket{\Psi_k(t)}
\end{eqnarray}
The formal solution can be expressed as
\begin{eqnarray}
\ket{\Psi_k(t)}=U_k(t,t_0)\ket{\Psi_k(t_0)}
\end{eqnarray}
\begin{figure}
\includegraphics[scale=.4]{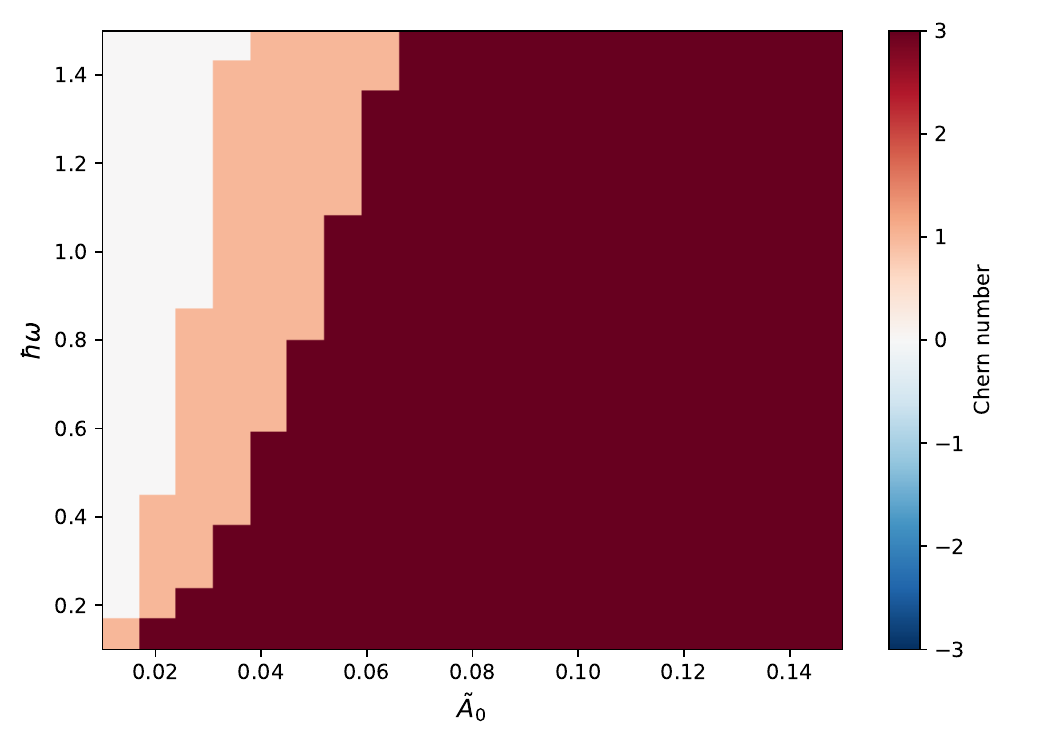}
\includegraphics[scale=.4]{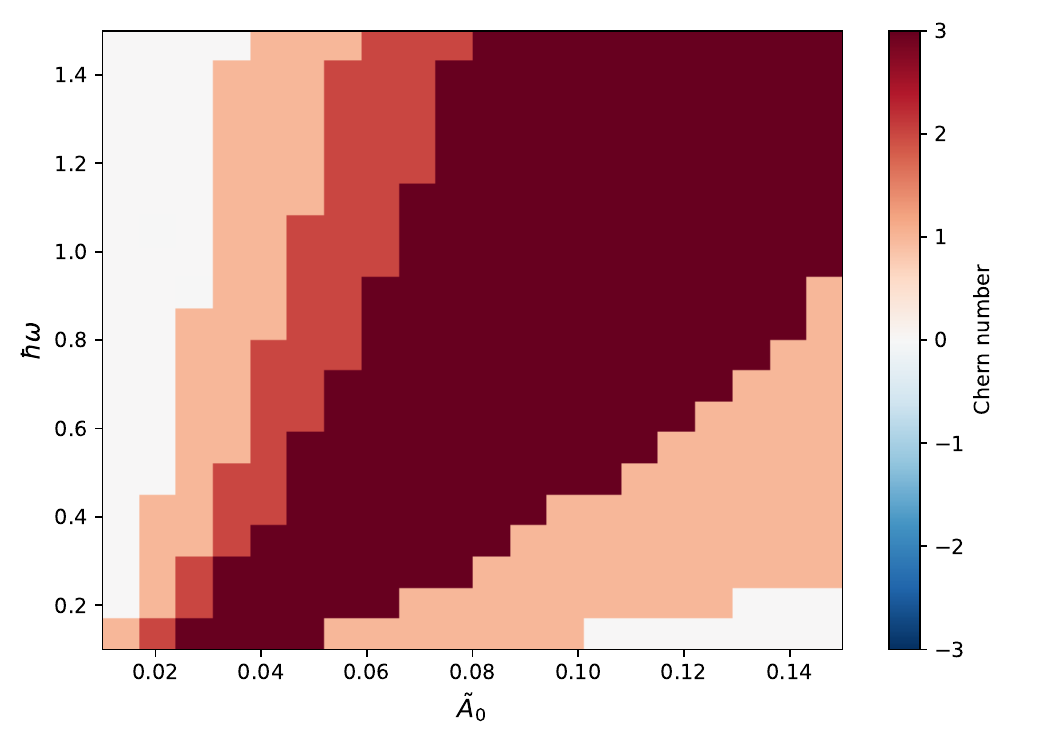}
\includegraphics[scale=.4]{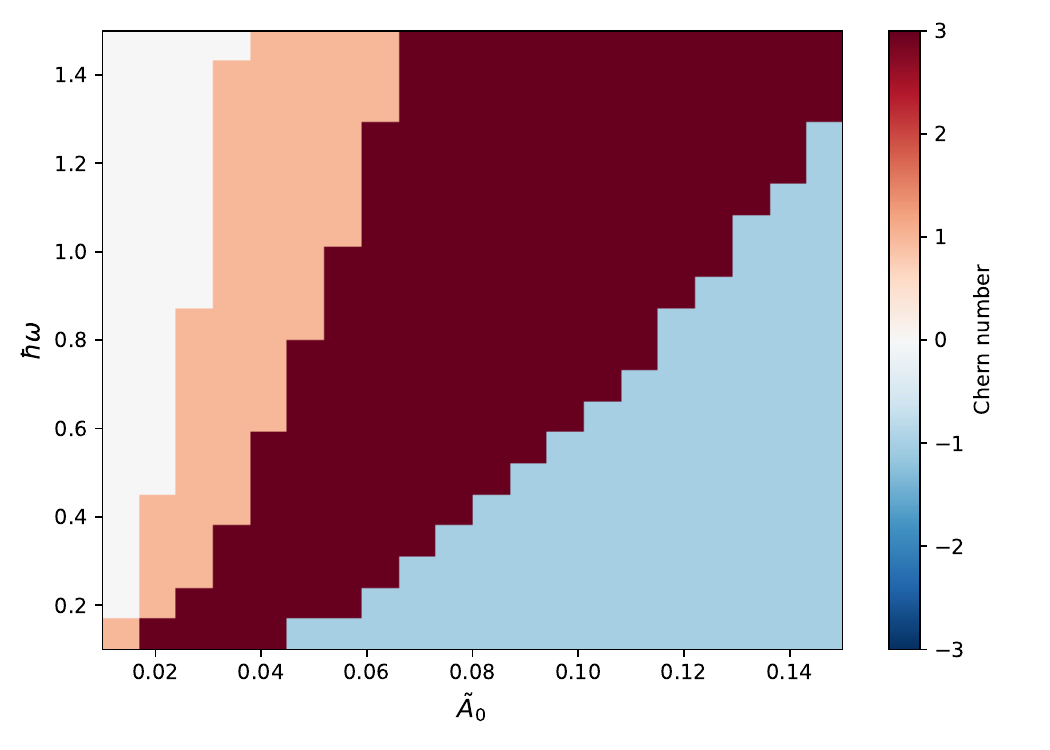}
\caption{The Chern number phase diagram of as a function of photon energy $\hbar \omega$ (in eV) and the dimensionless amplitude $\tilde{A}_0$ is shown. In the upper panel, the parameters are $\alpha_R=0$, $\alpha_D=0$, $\beta=0.5$ eV ${\AA}$. In the middle panel, the parameters are $\alpha_R=1$ eV ${\AA}$, $\alpha_D=0$, and $\beta=0.5$ eV ${\AA}^3$. In the lower panel, the parameters are $\alpha_R=0$, $\alpha_D=1$ eV ${\AA}$, and $\beta=0.5$ eV ${\AA}^3$. In all panels, the mass term is fixed at $\Delta_m=0.1$ eV.}
\label{phase-diag}
\end{figure}
where the time evolution operator is defined as
\begin{eqnarray}
U_k(t,t_0)&=&\mathcal{T}exp[-\frac{i}{\hbar}\int^t_{t_0}H_S(\vec{k},t')dt']
\end{eqnarray}
The stroboscopic evolution over one time period can be written as,
\begin{eqnarray}
U_k(t_0+T,t_0)&=&exp[-\frac{i}{\hbar}H_F(k)T]
\end{eqnarray}
which define the Floquet Hamiltonian $H_F(k)$. The eigenvalue problem $H_F(k)\ket{u_{k\alpha}}=\epsilon_{k\alpha}|\ket{u_{k\alpha}}$ yields quasienergies $\epsilon_{k\alpha}$, which are defined modulo $\hbar \omega$. In the off-resonant regime, the photon energy $\hbar \omega$ is much larger than all energy scales of the system (such as hopping, spin-orbit coupling), ensuring that real electronic transitions are suppressed. The periodic drive thus induces only virtual photon processes, which effectively renormalize the static band. Expanding the periodic Hamiltonian in Fourier components,
\begin{align}
H_S(\mathbf{k},t) &= \sum_{m} \mathcal{V}_m(\mathbf{k})\, e^{i m \omega t}, \\
\mathcal{V}_m(\mathbf{k}) &= \frac{1}{T} \int_0^T dt\, H_S(\mathbf{k},t)\, e^{-i m \omega t}.
\end{align}
the Floquet Hamiltonian can be approximated using the Magnus expansion as
\begin{eqnarray}
H_F(\mathbf{k})=H_0(\mathbf{k})+\sum_{m>1}\frac{1}{m\hbar \omega}[\mathcal{V}_{-m}(\mathbf{k}),\mathcal{V}_{+m}(\mathbf{k})]+\mathcal{O}(\frac{1}{\omega^2})\nonumber\\
\end{eqnarray}
The first term $H_0$ represents the time-averaged of static Hamiltonian. The second term arises from a sequence of absorb a photon of $+m$ and emit $-m$ (or vice versa). It is purely virtual process and generate a new term scales as $A^2_0/\omega$. The Higher order terms decay rapidly and are often negligible.
\begin{figure}
\includegraphics[scale=.4]{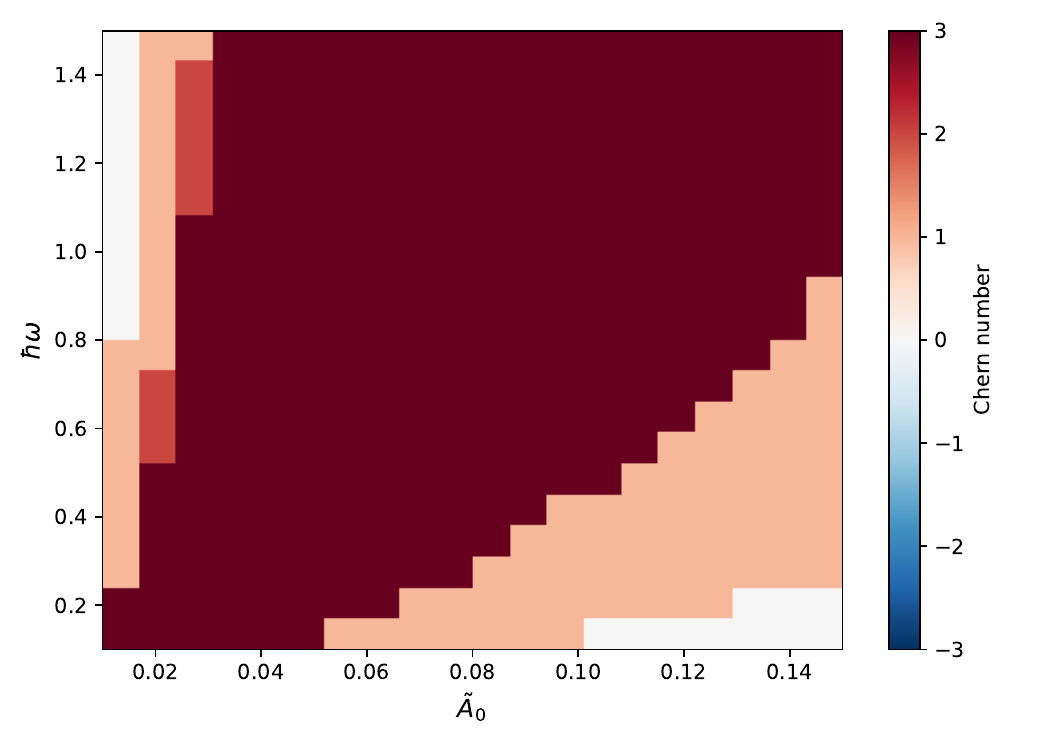}
\caption{The Chern number phase diagram as a function of photon energy $\hbar \omega$ (in eV) and the dimensionless amplitude $\tilde{A}_0$ is shown. The parameters used are $\alpha_R = 1~\text{eV},\text{\AA}$, $\alpha_D = 0$, and $\beta = 0.5~\text{eV},\text{\AA}^3$, with the mass term fixed at $\Delta_m = 0.1~\text{eV}$. The simulation is performed in the vicinity of the $\Gamma$ point, as described by Eq.~(\ref{linear-cubic-hamil})}
\label{gamma}
\end{figure}
Fig.(\ref{phase-diag}) presents the topological phase diagrams for different types of spin–orbit coupling. The upper panel corresponds to a system with pure cubic Rashba coupling, the middle panel includes both cubic and linear Rashba couplings, and the lower panel shows the case with cubic Rashba and linear Dresselhaus couplings. It is evident that the inclusion of linear Rashba coupling confines the topological phase to a bounded region in the parameter space spanned by the dimensionless drive amplitude $\tilde{A}_0$ and the photon energy $\hbar \omega$. In contrast, the presence of linear Dresselhaus coupling does not impose such a restriction, allowing the topological phase to persist over a broader range of parameters. In Fig.~(\ref{gamma}), we present the Chern number phase diagram obtained by considering the vicinity of the $\Gamma$ point, as described by Eq.~(\ref{linear-cubic-hamil}). The Chern numbers are evaluated numerically using the gauge-invariant Fukui–Hatsugai–Suzuki method\cite{Fukui-05}.

The emergence of different topological phases is associated with the merging and splitting of Dirac points at high-symmetry points in the Brillouin zone. Similar behavior, involving the merging of Dirac points in graphene and higher-order topological insulators under high-frequency driving, has also been discussed in Ref.\cite{Wan-PRB24},\cite{Delplace-PRB13}. This comparison highlights the distinct role of the linear Rashba term in stabilizing the Floquet-induced topological phase and demonstrates that the specific form of spin–orbit interaction plays a crucial role in shaping the topological phase diagram.

\section{Energy Balance and Dissipation Pathway}

The total Hamiltonian of the system can be written as
\begin{eqnarray}
H = H_S(t) + H_B + H_{SB},
\end{eqnarray}
where $H_S(t) = H_S(t+T)$ describes the periodically driven system, $H_B$ denotes the bath, and $H_{SB}$ represents the system–bath coupling Hamiltonian. 

We emphasize that the high-frequency (Floquet–Magnus) expansion employed in this work applies strictly to the coherent dynamics of the isolated system $H_S(t)$, describing the short-time behavior before energy dissipation becomes significant. The main effect of the coupling Hamiltonian $H_{SB}$ is to introduce damping, and the discussion below is intended solely as a qualitative illustration of experimental energy scales and relaxation rates rather than a microscopic open-system treatment.

In realistic settings, coupling to the environment allows the driven system to evolve toward a nonequilibrium steady state under continuous energy injection by the high-frequency drive. The characteristic timescale for reaching this steady state is $\Gamma^{-1}$, where $\Gamma$ represents the effective damping rate or energy scale associated with system–bath relaxation processes~\cite{Arakawa-PRB24,Arakawa-PRB25}. The average power absorbed per unit area can be estimated as
\begin{eqnarray}
P_{\mathrm{abs}} \approx \frac{1}{2}\sigma(\omega)E_0^2,
\end{eqnarray}
where $\sigma(\omega)$ is the real part of the optical conductivity. For a typical two-dimensional material with $\sigma(\omega) \approx 10^{-4}\,\Omega^{-1}$ and $E_0 = 10^7\,\mathrm{V/m}$, we obtain $P_{\mathrm{abs}} \approx 5\times10^9\,\mathrm{W/m^2}$. For photon energy $\hbar\omega \sim 1\,\mathrm{eV}$, the number of photons absorbed per unit area per second is 
$N_{\mathrm{photon/area}} = P_{\mathrm{abs}} / \hbar\omega \approx 10^{28}\,\mathrm{photons/(m^2\,s)}$. 
Taking an electron density $n_e = 10^{16}\,\mathrm{m^{-2}}$, the photon absorption rate per electron is 
$N_{\mathrm{photon/electron}} \approx 10^{12}\,\mathrm{s^{-1}}$. 
Since $N_{\mathrm{photon/electron}} \lesssim \Gamma \,(\sim 10^{12} - 10^{13}\,\mathrm{s^{-1}})$, the electrons can dissipate energy as fast as it is absorbed, maintaining a nonequilibrium steady state.

We note that the “bath” here is considered in a \textit{phenomenological Büttiker sense}—as an effective energy reservoir that enables relaxation and steady-state balance—rather than as a microscopic, Markovian quantum environment described by a Lindblad master equation. A rigorous treatment of dissipation in periodically driven systems would require a Floquet–Lindblad or Keldysh-type approach~\cite{Nuske-PRR20}, which lies beyond the scope of the present study. The present results thus remain fully within the coherent high-frequency expansion framework, and the remarks on dissipation and damping are included solely as a qualitative guide to experimental feasibility.

\section{Validity of the Fermi-Dirac Distribution in the Off-Resonant Regime}
For a periodically driven system coupled to the thermal bath, the transition rate from floquet state $\alpha$ to $\beta$ is given by a Fermi-Golden Rule expression\cite{Chamon-PRB15}
\begin{equation}
W_{\alpha\to\beta} = \sum_{n\in\mathbb{Z}} 
\Gamma(\Delta_{\alpha\beta}+n\hbar\omega)\,|S_{\beta\alpha}^{(n)}|^2,
\qquad \Delta_{\alpha\beta} = \epsilon_\alpha - \epsilon_\beta,
\end{equation}
where $S_{\beta\alpha}^{(n)}$ are the $n$-th Fourier components of the coupling matrix element and $\Gamma(\Omega)$ is the bath spectral density at frequency $\Omega$. For a thermal bath the Kubo-Martin-Schwinger condition holds, $\Gamma(-\Omega) = e^{-\beta\Omega}\Gamma(\Omega)$, with $\beta=1/k_BT$. If $n=0$ channel dominate (i.e., no net absorption or emission of photon), then,
\begin{eqnarray}
W_{\alpha\to \beta}\approx \Gamma(\Delta_{\alpha \beta})|S^{(0)}_{\beta\alpha}|^2\nonumber\\
W_{\beta\to \alpha}\approx \Gamma(\Delta_{\beta \alpha})|S^{(0)}_{\alpha\beta}|^2
\end{eqnarray}
So,the ratio of transition rates satisfies detailed balance:
\begin{equation}
\frac{W_{\alpha\to\beta}}{W_{\beta\to\alpha}} \approx e^{\beta(\epsilon_\alpha-\epsilon_\beta)}.
\end{equation}
This yields the stationary solution
\begin{equation}
p_\alpha^{\rm ss} = \frac{1}{e^{\beta(\epsilon_\alpha-\mu)}+1},
\end{equation}
which is the Fermi–Dirac distribution of Eq.~(\ref{an})\cite{Yan-PRL16}. The dominance of the $n=0$ sideband is guaranteed when
\begin{equation}
\frac{\sum_{|n|\ge 1} |S_{\beta\alpha}^{(n)}|^2\, \Gamma(\Delta_{\alpha\beta}+n\hbar\omega)}
{|S_{\beta\alpha}^{(0)}|^2 \,\Gamma(\Delta_{\alpha\beta})} \ll 1.
\end{equation}
In practice, two conditions ensure this suppression:  
(i) small dimensionless drive amplitude $\tilde A_0 = eA_0a/\hbar \ll 1$, so that $|S^{(n)}|$ rapidly decrease with $|n|$,  
and (ii) a bath cutoff frequency $\omega_c$ well below the drive frequency, $\hbar\omega \gg \hbar\omega_c$, such that $\Gamma(\Delta+n\hbar\omega)$ is negligible for $n\neq 0$.  

Thus, the Fermi–Dirac form in Eq.~(\ref{an}) is valid in the parameter regime
\begin{equation}
\boxed{\;\;\tilde A_0 \ll 1,\qquad \hbar\omega \gg \max\{W,\;\hbar\omega_c\}\;\;}.
\end{equation}
Outside this regime, photon-assisted transitions ($n\neq 0$) may populate high-energy states, and the steady state need not be thermal. Indeed, the experimental observations in Refs.\cite{He-Nano23},\cite{Dou-Small} correspond to such regimes and are consistent with deviations from the simple Fermi–Dirac description.

\end{document}